\documentclass[prd, preprintnumbers, nofootinbib, notitlepage, twocolumn]{revtex4}
\pdfoutput=1
\usepackage{graphicx}
\usepackage{color}
\usepackage{mathrsfs}
\usepackage{amsmath}
\usepackage{hyperref}
\newcommand{\pprime}{{\prime\prime}}

\newcommand{\sgn}{\mathop{\mathrm{sgn}}}

\newcommand{\be}{\begin{equation}}
\newcommand{\ee}{\end{equation}}

\begin{document}
\title{Validity of the kink approximation to the tunneling action}
\author{Koushik Dutta}
\author{Cecelie Hector}
\author{Thomas Konstandin}
\author{Pascal M.~Vaudrevange}
\author{Alexander Westphal}
\affiliation{DESY, Notkestrasse 85, 22607 Hamburg, Germany}
\preprint{DESY-12-025}

\begin{abstract}
  Coleman tunneling in a general scalar potential with two
  non-degenerate minima is known to have an approximation in terms of
  a piecewise linear triangular-shaped potential with sharp 'kinks' at
  the place of the local minima. This approximate potential has a
  regime where the existence of the bounce solution needs the scalar
  field to 'wait' for some amount of Euclidean time at one of the
  'kinks'. We discuss under which conditions a kink approximation of locally
  smooth 'cap' regions provides a good estimate for the bounce
  action.
\end{abstract}

\date{\today}
\maketitle
\section{Introduction}
A semi-classical approach to quantum tunneling processes in field
theory has been presented in a series of pioneering papers
\cite{Coleman:1977py}, \cite{Callan:1977pt}. The role of gravity in
the process of tunneling was subsequently considered in
\cite{Coleman:1980aw}. The authors presented a scheme for calculating
tunneling amplitudes for transitions from false to true vacua. The
calculation involves the evaluation of the Euclidean action of the
bounce solution to the imaginary-time equations of motion. The
existence of the bounce solution was proven in generality in
\cite{Coleman:1977py}. For almost degenerate vacuum energy, the
thin-wall approximation can be used to calculate the tunneling
amplitude without having to compute the bounce solution.

Given the fact that a bounce exists, a necessary and sufficient
condition in this scheme for the false vacuum to be unstable and
tunneling to occur is the existence of a single negative eigenvalue of
the operator $\delta^2 S$, the second variational derivative of the
Euclidean action evaluated at the bounce,
\cite{Coleman:1987rm}. Various authors examined systems where the
tunneling rate may become zero due to the non-existence of a bounce solution. This can e.g.~happen through the appearance of singularities in multi-field setups including gravity
\cite{Cvetic:1994ya}, \cite{Johnson:2008vn} that can be seen as the limiting case of a bouncing path that is extremely stretched \cite{Yang:2009wz}, \cite{Aguirre:2009tp}. Other examples for systems without bounce solutions are given by potentials with intermediate vacua \cite{Brown:2010bc}, \cite{Brown:2011um}. 
Therefore, the decay process
of false vacua via tunneling in the semi-classical picture firstly
depends on the existence of a bounce under consistent approximations,
and secondly on having only one negative eigenvalue of $\delta^2 S$.

In this short note, we would like to examine the validity of the kink approximation in the single field setup, i.e.~of bounce solutions in piecewise linear
potentials that act as approximations to locally smooth potentials. It
is clear that violating the conditions set out (implicitly) in the
proof of \cite{Coleman:1977py} is difficult in any physically
realistic setup.  Still this does not guarantee that the two pictures
lead to quantitatively similar actions.

\begin{figure}
  \includegraphics[width=0.95\linewidth]{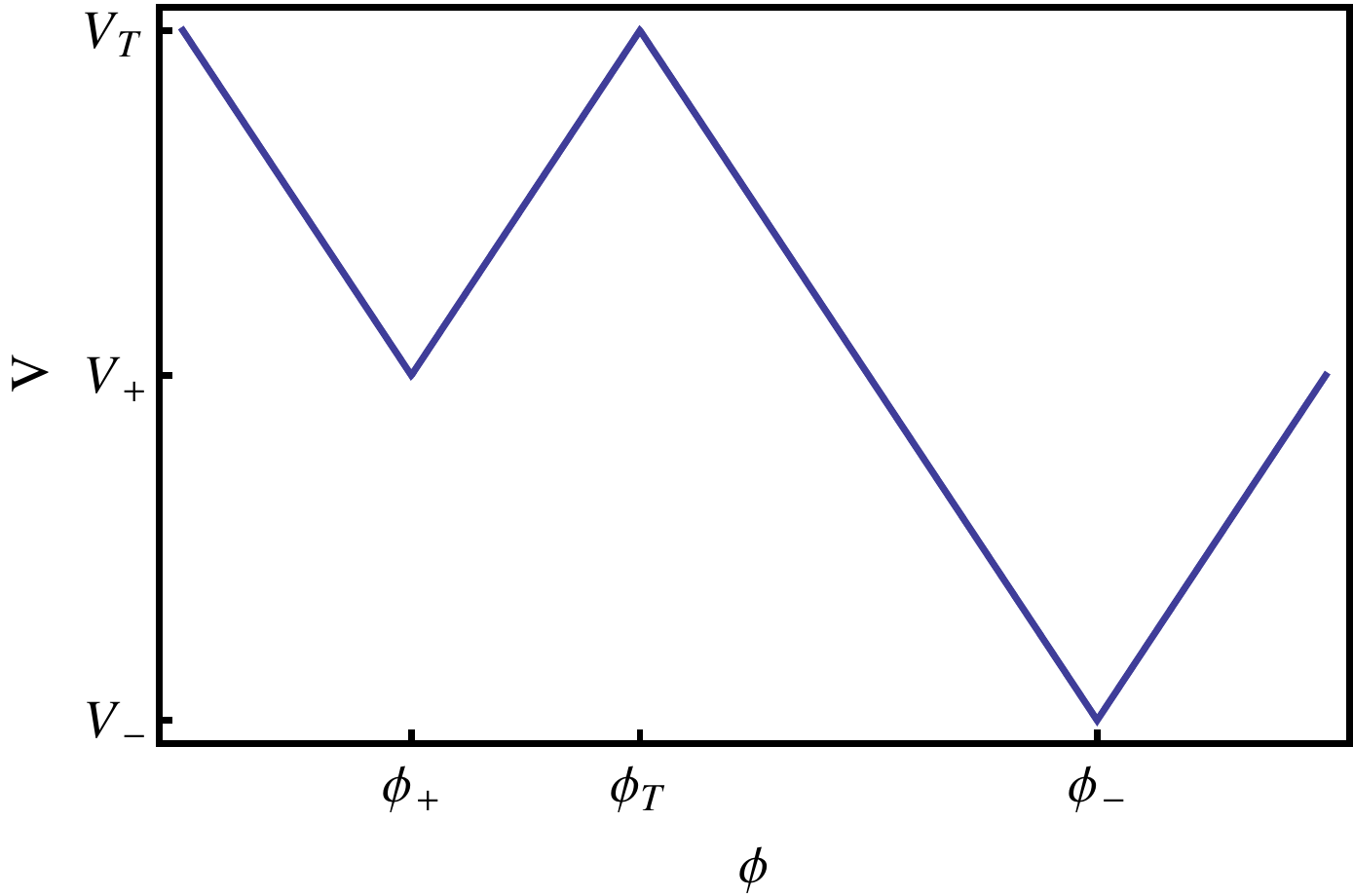}
  \caption{The piecewise linear potential first analyzed by
    \cite{Duncan:1992ai}. For $|\phi_+-\phi_T| > |\phi_--\phi_T|$, the
    bounce solution can only be found after modifying the initial 
conditions.}
  \label{fig:linear_linear}
\end{figure}

Consider an effective potential that has sharp minima and maxima
('kinks'), see Figure~\ref{fig:linear_linear}. Potentials of this shape can occur in the Randall-Sundrum scenario \cite{Randall:1999ee} and variants thereof \cite{Kaloper:1999sm, Brummer:2005sh}. Tunneling in this
setup was first discussed analytically in \cite{Duncan:1992ai}.  In
this case, for certain ranges of parameters, a consistent bounce
solution exists if we allow the field to rest for some amount of
Euclidean time at the true minimum.  With this field profile, it is
possible for the relevant friction term to die off sufficiently so
that the field can roll back up to the false vacuum.  Having the
field 'wait' in such a manner is only an approximation to the
physical situation in which the tip of the potential is replaced by a smooth
cap.



In the next Section we briefly review the original arguments for the
existence of bounce solutions in the single field setup. In Section III, we discuss the
tunneling in a piecewise linear potential and explain how smoothing of
the potential is important for certain parameter ranges of the
potential. This subsequently leads us to a condition on the smooth
potential for a meaningful approximation in terms of a piecewise
linear potential in Section IV. In Section V, we comment on the
applicability of the shooting algorithm in the smooth potential and we
conclude in Section VI.

\section{Existence of bounce solutions}

In his pioneering paper \cite{Coleman:1977py}, Coleman offered an
existence proof for the bounce solution. We briefly sketch this proof
before we specialize to piecewise linear potentials in the next
Section.
\begin{figure}
  \includegraphics[width=0.95\linewidth]{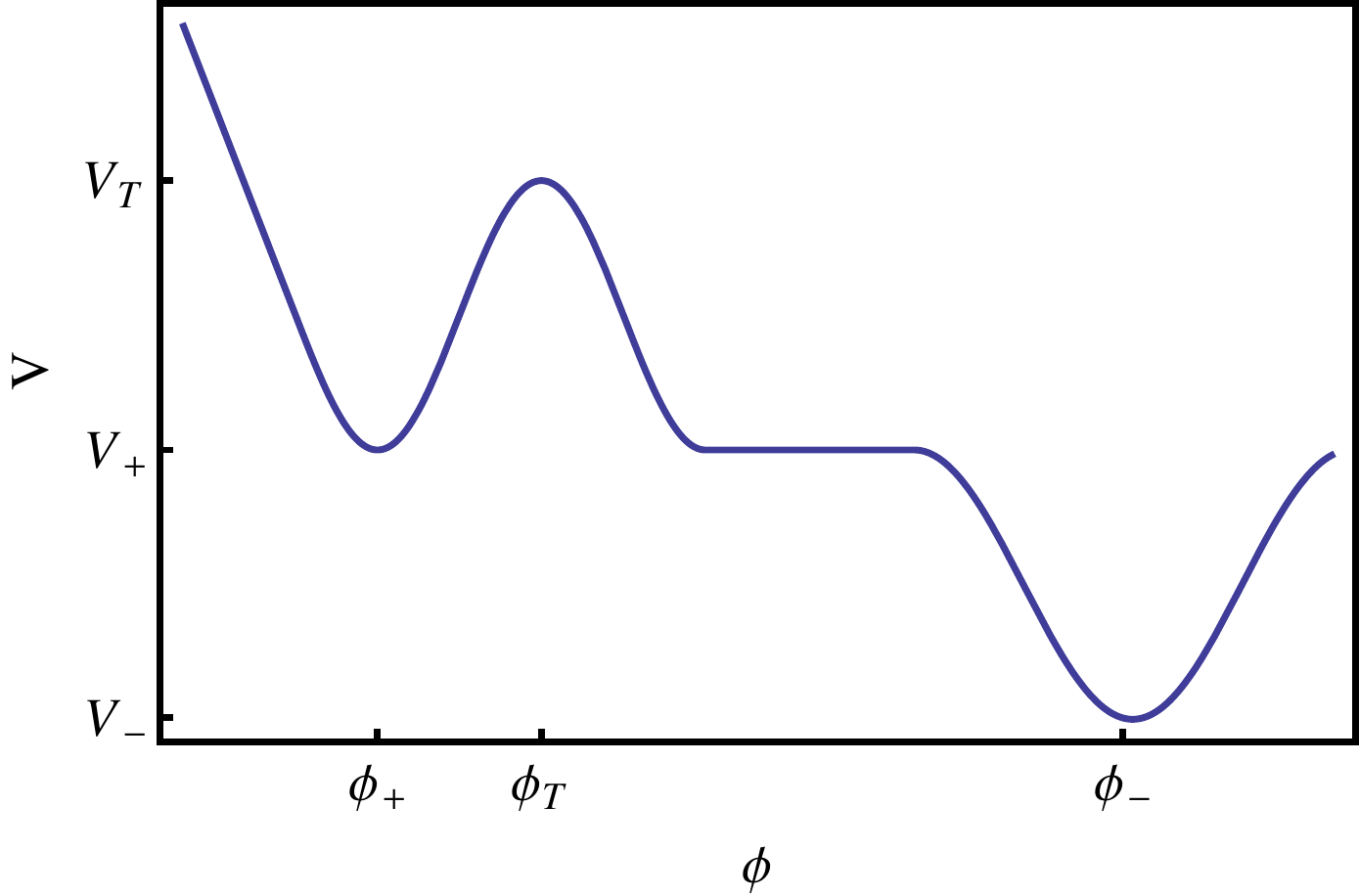}
  \caption{Schematic view of a potential with an existing bounce
    solution. The crucial part is that the true minimum is a smooth
    function of $\phi$, i.e.~$\partial_\phi V$ is continuous.  }
  \label{fig:bounce_exists}
\end{figure}

In the inverted potential (see Figure~\ref{fig:bounce_exists}), the
bounce solution $\phi_B$\footnote{The term bounce stems from the
  description of this process in QM, where the field sits in the false
  vacuum for $t\to-\infty$, reaches the true vacuum at some
  $-\infty<t_*<\infty$, and rolls back to the false vacuum for
  $t\to\infty$.}  is the solution to the equation of motion (in $3+1$
dimensional Euclidean space time)
\begin{eqnarray}\label{eq:existence:eom}
  \phi^\pprime + \frac{3}{r}\phi^\prime - \partial_{\phi} V&=&0\,,
\end{eqnarray}
where $\phi^\prime\equiv\partial_r\phi$ with the following
properties. At the center of the bubble, at $r=0$, the field sits with
zero velocity at position $\phi_0$ somewhere between $\phi_T$, the
location of the top of the potential barrier, and $\phi_-$, the
location of the true vacuum. For $r>0$, the field moves towards the
false vacuum $\phi_+$, reaching it with zero speed for $r\to\infty$.
Owing to the friction term in Eq.~\eqref{eq:existence:eom}, it is not
immediately clear that the field can ever reach $\phi_+$.

Inside of the bubble, starting ever closer to $\phi_-$, the field can
sit almost fixed at that position for a longer and longer time --
until friction dies off. Then, energy conservation makes the field
roll past $\phi_+$ as long as $\Delta V_-\ge\Delta V_+$ ($\phi_\pm$
being the location of the false and true vacuum respectively, and
$\Delta V_{\pm} = V_T - V_{\pm}$). To show that overshooting past
$\phi_+$ occurs for starting values $\phi_0$ close enough to $\phi_-$,
we need to be somewhat quantitative: For analytic potentials it is
possible to linearize the equation of motion close to the true vacuum
$\phi_-$, giving
\begin{eqnarray}\label{eq:existence:eom:linearized}
  \left(\partial_r^2+\frac{3}{r}\partial_r -\mu^2\right)(\phi-\phi_-)&=&0\,,
\end{eqnarray}
with $\mu^2 = V''(\phi_-)$. It is solved by
\begin{eqnarray}
  \phi(r)-\phi_-&=&2(\phi_0-\phi_-)\frac{I_1(\mu r)}{\mu r}\,,
\end{eqnarray}
where $I_1$ is the Bessel function of the first kind, see
\cite{Coleman:1977py}. Hence for $\phi_0$ ever closer to $\phi_-$, the
field can sit near $\phi_-$ for larger and larger $r$. Making the
initial displacement from $\phi_-$ sufficiently small, $r$ becomes
large enough such that the friction term effectively
disappears. Thus by energy conservation, $\phi$ can rush past
$\phi_+$.

On the other hand, starting far away from the top of the inverted
potential, the field does not have enough energy to climb up to
$\phi_+$. Thus, by continuity, there must be an initial value $\phi_0$
being $\phi_T<\phi_0<\phi_-$ such that the field reaches $\phi_+$ with
zero velocity.

\section{Piecewise linear potentials}

The existence of the bounce crucially depends on the
possibility of the field to spend arbitrarily long times arbitrarily
close to the true vacuum. In other words, if the equation of motion
\eqref{eq:existence:eom:linearized} takes on a different form, it is
not guaranteed that the field can spend enough time near the true
minimum for the friction term to die out. In particular, it is
intuitively clear that this is the case for piecewise linear
potentials, see Figure~\ref{fig:linear_linear}. In the following, we
discuss the tunneling solutions in detail for a piecewise linear
potential, pointing out several subtleties before we analyze the
transition to the smooth and regular potential where the kinks are
resolved by caps.

Tunneling in a piecewise linear potential
\begin{eqnarray}\label{piecewise_linear_potential}
  V&=&\left\{\begin{array}{cc}
  V_T+\lambda_+(\phi-\phi_T)\,,&\phi<\phi_T\\
  V_T-\lambda_-(\phi-\phi_T)\,,&\phi\ge\phi_T
  \end{array}
  \right.\,,
\end{eqnarray}
has been analyzed by \cite{Duncan:1992ai}, see
Figure~\ref{fig:linear_linear}. We present their analysis in a
slightly different form.

First of all, we set $\phi_T=V_T=0$ as shifts in the field and in the
zero point energy do not change the physics -- ignoring the effects of
gravity. Solving the equation of motion inside of the bubble
\begin{eqnarray}\label{eq:phi_i:eom}
  \phi_i^\pprime+\frac{3}{r}\phi_i^\prime+\lambda_-&=&0\,,
\end{eqnarray}
subject to $\phi(0)=\phi_0$, $\phi^\prime(0)=0$, we find
\begin{eqnarray} \label{inside_bounce}
  \phi_i&=&\phi_0-\frac{\lambda_-}{8}r^2\,.
\end{eqnarray}
Enforcing the matching condition $\phi_i(R_T)=0$ gives
\begin{eqnarray}
  R_T&=&2\sqrt{\frac{2\phi_0}{\lambda_-}}\,.
\end{eqnarray}
Solving the equation of motion on the outside of the bubble
\begin{eqnarray}
  \phi_o^\pprime+\frac{3}{r}\phi_o^\prime-\lambda_+&=&0\,,
\end{eqnarray}
subject to $\phi_o(R_T)=0$, $\phi_o^\prime(R_T)=\phi_i^\prime(R_T)$, we
find
\begin{eqnarray} \label{outside_bounce}
  \phi_o&=&\frac{(R_T^2-r^2)(R_T^2(\lambda_-+\lambda_+)-r^2\lambda_+)}{8r^2}\,.
\end{eqnarray}
On the outside, the field settles in the false vacuum at radius
$R_+>R_T$, for which $\phi_o^\prime(R_+)=0$
\begin{eqnarray}
  R_+&=&\left(1+\frac{\lambda_-}{\lambda_+}\right)^{1/4}R_T\,.
\end{eqnarray}
At this position, the field has the value
\begin{eqnarray}\label{eq:phi_o_max}
  \phi_o(R_+)&\equiv&\phi_+=-\phi_0\frac{\sqrt{1+c}-1}{\sqrt{1+c}+1}\,,
\end{eqnarray}
where $c=\lambda_-/\lambda_+$. As $\phi_0<\phi_-$, this gives a
restriction on the shape of the potential
\begin{equation}
  \alpha\equiv\frac{-\phi_+}{\phi_-}\le\frac{-\phi_+}{\phi_0}=\frac{\sqrt{1+c}-1}{\sqrt{1+c}+1}<1\,.
\end{equation}
This is equivalent to
\begin{eqnarray}
  \frac{\Delta V_+}{\Delta V_-}&\le&\frac{1}{2+c+2\sqrt{1+c}}\le\frac14\,.
\end{eqnarray}
In other words, the bounce solutions given by
Eq.~\eqref{inside_bounce} and Eq.~\eqref{outside_bounce} with initial
condition $\phi(0)=\phi_0$, $\phi^\prime(0)=0$ are only valid for the
parameter range $\alpha<1$ and $\Delta\equiv\Delta V_+/\Delta V_-<\frac{1}{4}$.

As was already pointed out by \cite{Duncan:1992ai}, outside of this
parameter range, the initial conditions need to be modified to find a
bounce solution. These modifications are explored in more detail
below. But before proceeding, let us take a look at the physical
meaning of these conditions. It corresponds to a potential profile where
the energy difference between the false and the true vacuum is large
(i.e.~{\it~not} thin-wall) and $|\phi_+| < |\phi_-|$. In this case, it
is always possible to find a field position $\phi_0 < \phi_-$ with
zero initial velocity such that the field can roll up the hill to
$\phi_+$.

On the other hand, if either $\Delta > 1/4$ for any value of $\alpha$,
or $\Delta < 1/4$ and $\alpha > 1$, it is not immediately clear that a
bounce solution exists for $\phi_0 \le \phi_-$. The physical picture
is as follows: 

For $\Delta > 1/4$ (and any value of $\alpha$), there is a smaller
energy difference between the true and the false minimum -- in the
extreme case, making the energy difference infinitesimally small for
$\Delta\to1$, the thin-wall limit. For almost degenerate minima, the
field would need to wait near the true minimum for the friction term
to fall off. But the linear potential makes it impossible for the
field to stay longer at that initial position.  Therefore, with a
small energy difference between the true and the false vacuum, the
field can not roll up the hill due to friction, and no solution exists
that reaches $\phi_+$.

For $\Delta < 1/4$ and $\alpha > 1$, although the field has a large
potential energy to start with, the non-vanishing friction term still
prevents it from climbing up the long shallow part to reach $\phi_+$.

Another example in which a large difference in potential energy does
not guarantee a bounce solution are scalar potentials with local
$\phi^4$ (or higher power) behavior
\begin{eqnarray}\label{eq:neg_quartic}
  V&=&-\frac{c_4}{4}(\phi - \phi_*)^4+\text{higher order}\,,\quad c_4>0\,.
\end{eqnarray}
\begin{figure}
  \includegraphics[width=0.95\linewidth]{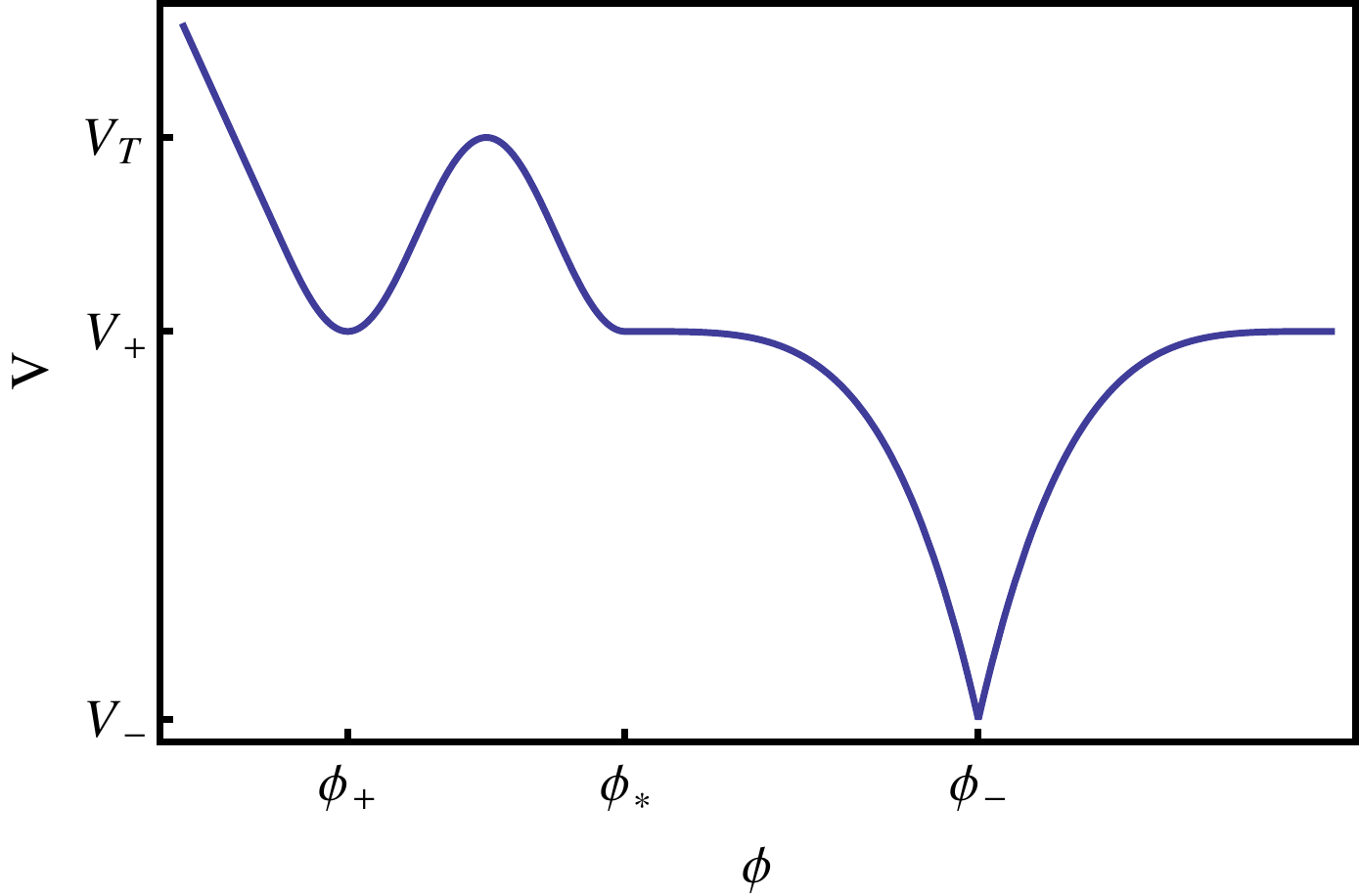}
  \caption{Shape of a true minimum for which no bounce exists. If
    the wings of the trough at $\phi_-$ are polynomials of at least
    order $4$, the bounce solution in the inverted potential cannot
    roll past the point $\phi_*$ on the left.}
  \label{fig:acute_minimum}
\end{figure}
For solutions to the equation of motion
\begin{eqnarray}
  \phi^\pprime+\frac{3}{r}\phi^\prime+ c_4(\phi - \phi_*)^3&=&0\,,
\end{eqnarray}
with initial conditions $\phi(0)=\phi_0$ and $\phi'(0)=0$, the field
reaches $\phi=\phi_*$ with zero speed~\cite{Dutta:2011fe}
independent from the release point $\phi_0$. If the $\phi^4$ behavior
ends with a kink (as depicted in Figure~\ref{fig:acute_minimum}) no
bounce solution exists.

It is important to note that the arguments illustrated in Section II
do not immediately hold here: there is no quadratic approximation of
the potential around the true minimum to get
Eq.~\eqref{eq:existence:eom:linearized}. In a linear potential,
putting the initial position ever closer to the true minimum $\phi_-$
does not force the field to spend an ever longer time there.

As \cite{Duncan:1992ai} pointed out, one thus needs to modify the
initial conditions. Inside the bubble, the field should be
artificially fixed at the true minimum $\phi_-$ until radius
$R_0$. This should be understood as an approximation. In particular,
this can be interpreted as mimicking the effect of removing the kink
and replacing it with a smooth cap. In this suitably capped potential,
the field can sit arbitrarily close to the true minimum and spend ever
longer time there. It is clear that the original argument of the
existence of bounce solution as outlined in Section II immediately
holds.

This waiting period can be realized by a change in the boundary
conditions as first done by \cite{Duncan:1992ai} for a piecewise
linear potential. The modified initial conditions become
$\phi_i(r)=\phi_-$, $\phi_i^\prime(r)=0$ for $0\le r\le R_0$, giving
\begin{eqnarray}
\label{eq:inner_sol}
  \phi_i(r)&=&\phi_--\frac{(r^2-R_0^2)^2}{8r^2}\lambda_-\,
\end{eqnarray}
for $ R_0 \leq r < R_T$. Outside of the bubble at $R_+>R_T$, the field
comes to rest in the false vacuum $\phi_o^\prime(R_+)=0$,
$\phi_o^\prime(R_+)=0$, giving
\begin{eqnarray}
  \phi_o(r)&=&\phi_++\frac{(r^2-R_+^2)^2}{8r^2}\lambda_+\,.
\end{eqnarray}
Now matching the two solutions $\phi_i(R_T)=0=\phi_o(R_T)$,
$\phi_i^\prime(R_T)=\phi_o^\prime(R_T)$ gives
\begin{eqnarray}
  R_0&=&\frac{\phi_-}{1-\sqrt{\Delta}}\sqrt{\frac{2}{V_-}}\sqrt{(1+\alpha)\left[\alpha-1+2\sqrt{\Delta}\right]}\,,\nonumber\\
  R_T&=&\frac{\phi_-}{1-\sqrt{\Delta}}\sqrt{\frac{2}{V_-}}(1+\alpha)\,,\\
  R_+&=&\frac{\phi_-}{1-\sqrt{\Delta}}\sqrt{\frac{2}{V_-}}\sqrt{(1+\alpha)\left[1+\alpha\left(\frac{2}{\sqrt{\Delta}}-1\right)\right]}\,,\nonumber
\end{eqnarray}
with $R_0<R_T<R_+$, $\alpha=-\frac{\phi_+}{\phi_-}$, $\Delta
V_\pm=V_T-V_\pm$ and $\Delta=\frac{\Delta V_+}{\Delta V_-}$. Note
that the condition that $R_0$ is real implies that $\alpha >
1-2\sqrt{\Delta}$.

The tunneling amplitude can then be computed as
\begin{eqnarray}
  B&=&B_a+B_b+B_c\,,
\end{eqnarray}
with 
\begin{eqnarray}
  B_a&=&2\pi^2\int_0^{R_0}\!dr\,r^3\left(-\lambda_-\phi_--\lambda_+\phi_+\right)\,,\\
  B_b&=&2\pi^2\int_{R_0}^{R_T}\!dr\,r^3\left(\frac{1}{2}\phi_L^{\prime2}-\lambda_-\phi_L-\lambda_+\phi_+\right)\,,\\
  B_c&=&2\pi^2\int_{R_T}^{R_+}\!dr\,r^3\left(\frac{1}{2}\phi_R^{\prime2}+\lambda_+\phi_R-\lambda_+\phi_+\right)\,,
\end{eqnarray}
giving
\begin{eqnarray}\label{eq:zero_speed_amplitude}
  B &=&\frac{2\pi^2}{3}\frac{\phi_-^4}{\Delta V_-}\frac{(1+\alpha)^3\left((\alpha-3)\sqrt{\Delta}+1-3\alpha\right)}{(\sqrt{\Delta}-1)^3}\,.
\end{eqnarray}

For certain choices of parameters of a piecewise linear potential, we
just saw that the bounce solutions exist only when we keep the field
artificially fixed at the true minimum. Holding it there for a
sufficiently long time, the damping term becomes small enough such
that the field can reach the false vacuum with zero velocity. This
should be thought of as an approximation to the physical situation of
smoothing the tip with a cap.

\section{Caps versus kinks}

In this Section, we discuss in which cases replacing the tip of a
piecewise linear potential with a smooth cap can be well approximated
by keeping the field artificially fixed at the minimum as outlined in
the previous Section. This means that the original argument by Coleman guarantees the existence of a bounce. Here we focus on the error in the tunneling action which is introduced by the kink approximation.

\begin{figure}
  \includegraphics[width=0.95\linewidth]{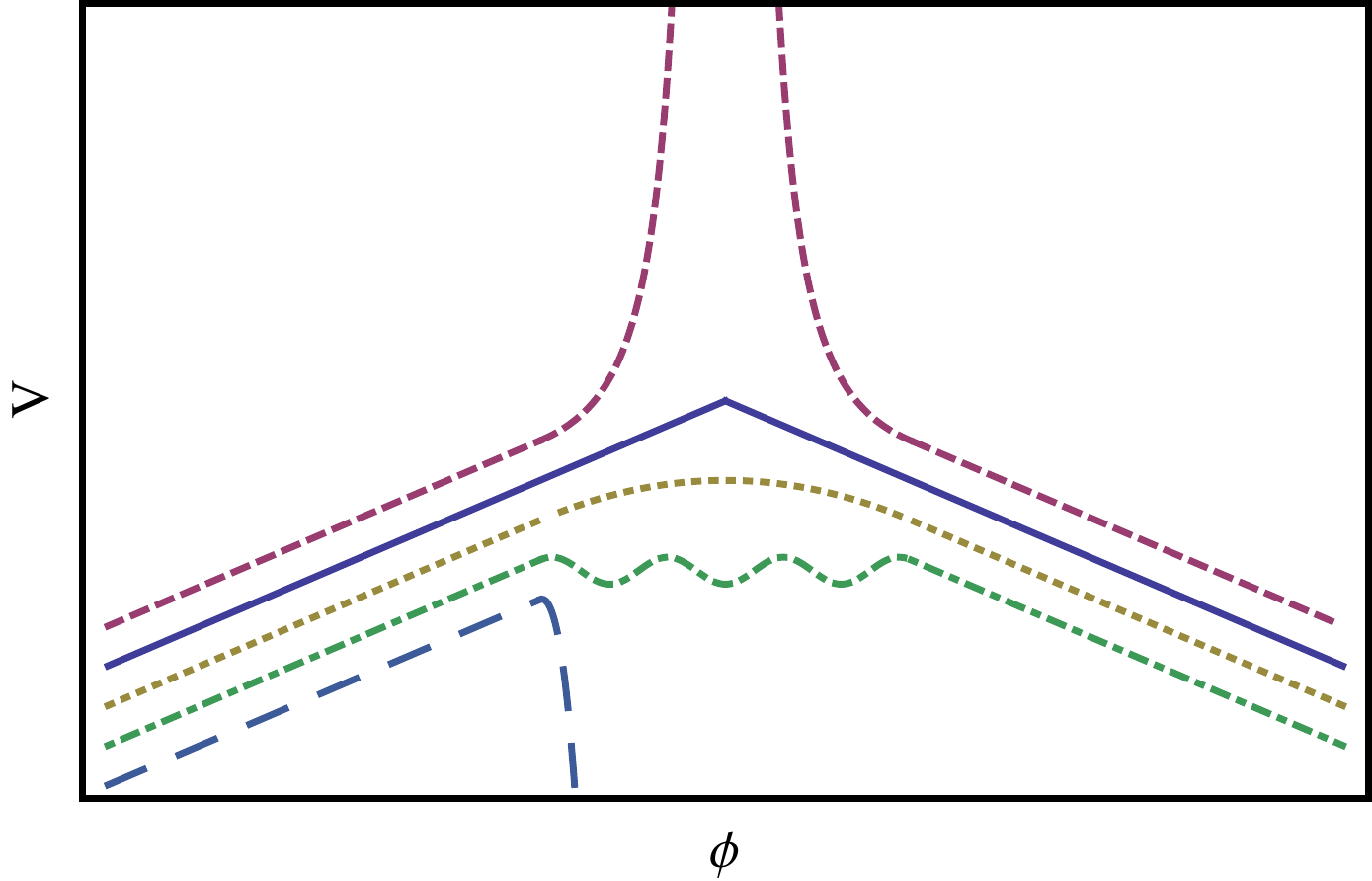}
  \caption{ The piecewise linear potential and different regular potentials
    in the cap region. The three potentials on the bottom pass the
    criterion (\ref{eq:cond_1a}) while the potential on top can lead
    to large deviations in the bounce action.}
  \label{fig:cap_examples}
\end{figure}

Suppose that a piecewise linear potential is obtained as the limit of
a regular smooth potential. The scale $\delta \phi$ on which the kink
is resolved in the regular potential serves as expansion parameter
$\delta \phi \ll |\phi_T-\phi_-|$. Apparently, the bounce actions can
be very different, if the smooth potential varies strongly in the cap
region. For example, if the potential has a large positive spike, the
bounce solution can leave the cap region with a sizable velocity that
can alter the bounce action significantly. Hence we demand that the
potential in the cap does not differ too much from the corresponding
value $V_-$ in the kink potential at least up to the first local
minimum in the regularized potential
\be
\label{eq:cond_1a}
|V_-^{\rm{cap}} - V^{\rm{kink}}_-| \lesssim  \lambda_- \delta \phi.
\ee
Still, the potential can vary strongly in the cap region in the sense
that its derivative does not need to be small. Some examples are given
in Figure~\ref{fig:cap_examples}.

Now, consider the bounce solutions for the kink potential and the cap
potential outside the cap region. Even for large $r$ these two
solutions only coincide approximately. It might well be that one of
the two solutions passes a given point in the potential a little bit
later but compensates by a slightly smaller velocity. The former
effect leads to a reduced friction that is compensated by the latter
effect.

Even a small difference between the two bounces can have a large
effect on the action. To see this, consider a piecewise linear
potential with two slightly different kink positions $\phi_-$ but same
slope $\lambda_-$. Suppose that we would arrange this shift in
$\phi_-$ such that the bounce solution of this modified kink potential
and the bounce of the regular potential coincide. The potential with
the more remote kink position $\phi_-$ has a smaller $R_0$.  According
to (\ref{eq:zero_speed_amplitude}), a shift $\Delta \phi_-$ leads
generically to a change in the bounce action of order $\delta B /B
\sim \Delta \phi_- / \phi_-$ and hence be small. Only in the thin wall
regime where $\Delta \simeq 1$ this change can be large and of order
\be
\frac{\delta B}{B} \simeq \frac32 
\frac{1}{\sqrt{\Delta} - 1} \frac{\Delta \phi_-}{\phi_-}.
\ee
Fortunately, $\Delta \phi_-$ cannot be larger than $\delta \phi$. We
prove this by contradiction. The solution to the field equations of
motion are (see Eq.~(\ref{eq:inner_sol}))
\begin{eqnarray}
\label{eq:inner_sol_shifted}
  \phi_i(r)&=&\phi_- + \Delta \phi_- 
- \frac{(r^2-(R_0 + \delta R_0)^2)^2}{8r^2}\lambda_- \ .
\end{eqnarray}
Now assume $\delta \phi \ll \Delta \phi_- \ll \phi_-$: the field 
leaves the cap at
\be
r^2 - (R_0 + \delta R_0)^2 \simeq \sqrt{8 \Delta \phi_- (R_0 + \delta R_0)^2 / \lambda_-}
\ee
with velocity 
\be
\phi_i' \simeq \sqrt{2 \Delta \phi_- \lambda_-}.
\ee
However, due to energy conservation and the condition
(\ref{eq:cond_1a}), the energy at the border of the cap cannot exceed
the potential energy. This implies
\be
 \phi_i^\prime \lesssim \sqrt{2  \lambda_- \delta \phi},
\ee
and hence $\Delta \phi_- \lesssim \delta \phi$ such that
\be
\label{eq:constr_dBB}
\frac{\delta B}{B} \lesssim \frac32 
\frac{1}{\sqrt{\Delta} - 1} \frac{\delta \phi}{\phi_-} \,.
\ee
Thus we demand
\be
\label{eq:cond_1b}
\sqrt{\Delta} - 1 \gg \frac{\delta \phi}{\phi_-},
\ee
in order to obtain accurate results for the bounce action in the
kink approximation.

Even though our reasoning above seems very conservative, no better
upper bound on the variation of the action exists in the thin-wall
regime. The boundary of the relation (\ref{eq:cond_1b}) is equivalent
to
\be
| V_+ - V_-| \simeq \lambda_- \delta \phi\,,
\ee
but in this case the potential difference between the true and false
vacua in the regular capped potential can be very different from that in
the kink potential. This can lead to grossly different bounce actions.
For this special situation, the constraint in \eqref{eq:constr_dBB} is
saturated.

Combining the criteria \eqref{eq:cond_1a} and \eqref{eq:cond_1b}, our
two conditions on the kink approximation read
\be\label{kinkcond}
|V_-^{\rm cap} - V^{\rm kink}_-| 
\lesssim \lambda_- \delta \phi 
\ll | V^{\rm kink}_+ -  V^{\rm kink}_-| \, .
\ee
In general, the kink approximation turns out to be very robust, as eq.~\eqref{kinkcond} only limits strong variations of the scalar potential within the smooth ' cap' region. The exception to the general case is given by the thin-wall limit. There, the tunneling action depends sensitively on the bubble wall tension and the vacuum energy difference. Hence, in this regime the thin-wall approximation is more appropriate. The kink approximation agrees with the thin-wall limit if it keeps constant the bubble wall tension and the vacuum energy difference.

\section{Exotic caps}

In the preceding Sections, we demonstrated that under most
circumstances, smoothing the kink in a piecewise linear potential is
equivalent to holding the field fixed at some radius $R_0$. This
statement depends crucially on the choice of cap that replaces the
kink. The shape of the cap must be such that the field can spend an
arbitrarily long time $R$ close to the true vacuum to allow the
friction term to get sufficiently small. This time $R$ is
approximately (assuming that friction dominates over acceleration in
the equation of motion)
\begin{eqnarray}
  \frac{3}{r}\frac{d \phi}{dr}\approx V'(\phi)&\Rightarrow&\int_{\phi_{0}}^{\phi(R)}\!\!\frac{d\phi}{V'(\phi)}=\frac{1}{6}R^2\,.
\end{eqnarray}
To spend an arbitrary and potentially infinite amount of time near the
cap, the integral must diverge in the limit $\phi_0\to
\phi_-$. Certainly this is true for analytic potentials with a finite
mass $|V''(\phi_-)|<\infty$. For example, taking $V(\phi)=\frac12
m^2\phi^2$, it is clear that the integrand becomes
$1/V'(\phi)=1/m^2\phi$ which is logarithmically divergent.

However, potentials such that the integral is finite in the limit
$\phi_0\to\phi_-$ do also exist. One class of examples are potentials
of the form $V\propto|\phi-\phi_-|^\alpha$, with $1<\alpha<2$, which
we shall now analyze in more detail. The derivative of the potential
at the minimum is $V'(\phi_*)=0$ in the limit $\phi_*\to\phi_-$ both
from the left and from the right.

To set up the full picture, let us assume a piecewise potential, where
we examine the area around the cap. We assume that the other piecewise
parts of the potential contain at least one other local minimum. The
complete bounce is given by matching the solutions in each part of the
potential. We solve the equations of motion for the bounce in the part
of the potential describing the cap where $V(\phi)=\lambda
|\phi-\phi_-|^\alpha$, $1<\alpha<2$. The equation of motion reads
\begin{eqnarray}
  \phi^\pprime+\frac{3}{r}\phi^\prime-\lambda 
  \alpha \sgn(\phi-\phi_-) |\phi - \phi_-|^{\alpha-1}&=&0\,,\label{eq:eom:alpha}
\end{eqnarray}
subject to the boundary conditions $\phi(0)=\phi_0, \, \phi^\prime(0)=0$.
Even in the limit $\phi_0 \to \phi_-$, the solution spends only a
finite amount of time in the cap region and is given by
\begin{eqnarray}
  \phi(r)&=& \phi_- + \left(\frac{4\alpha\lambda (\alpha-2)^2}{3-\alpha}
\right)^{\frac{1}{2-\alpha}} r^{\frac{2}{2-\alpha}}\,.\label{eq:dome1}
\end{eqnarray}
Hence, even though the field starts off from the extremum where there
is no force ($V'(\phi_-)=0$), it begins rolling away in finite time.
However, this does not imply that there is no bounce. The field can
wait for some time at $\phi_-$ and then still leave the cap in finite
time (friction is even less important than in the previous case).
This solution can be obtained by using the boundary conditions
$\phi(R_0)=\phi_0$ and $\phi'(R_0)=0$, and sending the release point
subsequently to $\phi_-$. If the waiting period $R_0$ is chosen
appropriately, the field reaches the false vacuum, thus constituting a
bounce solution. 

Numerically, the bounce solution is often determined using the
shooting algorithm~\cite{Coleman:1977py}. In this case, a release
position for the field is chosen and the corresponding initial value
problem using the equation of motion is solved. The release point is
then varied until the correct boundary conditions of the bounce
solution in the false vacuum are fulfilled. Obviously, a bounce of
the kind described above cannot be found with the conventional shooting
algorithm.

This situation might be rather surprising, since the potential is
smooth and even differentiable everywhere. This special situation
arises because the equation of motion for the bounce in the potential
\eqref{eq:eom:alpha} with $\alpha<2$ does not fulfill the Lipschitz
condition. Hence the Picard-Lindel\"of theorem does not hold and a
solution to the initial value problem is not necessarily unique. This
kind of situation is well known in the philosophy of science community
\cite{Norton:2008} in the context of classical mechanics. Fortunately,
here we need not concern ourselves with the initial value problem. We
are interested in computing the tunneling amplitude, and the bounce
solution with the usual boundary conditions is indeed unique. So for
this kind of smooth and regular potential, one can be in the situation that the kink
approximation describes the tunneling process quite well, while the
common shooting algorithm fails.

\section{Conclusions}

We have addressed the range of validity the kink approximation to Coleman tunneling in a smooth potential. Tunneling in a piecewise linear potential with kings was first
discussed analytically in \cite{Duncan:1992ai}. For certain ranges of
parameters, a consistent bounce solution exists only if the field can
rest for some amount of Euclidean time at the true vacuum. With this
field profile, it is possible for the relevant friction term to die
off sufficiently so that the field can roll back up to the false
vacuum. Having the field 'wait' in such a manner is only an
approximation to the full bounce solution where the tip represents a
smooth cap of a regular potential.

We found that replacing a regular smooth potential by its piecewise
linear approximation is a very robust procedure. A sufficient
criterion for the bounce action of the kink potential to yield
accurate results is given by 
\be\label{capcond}
|V_-^{\rm{cap}} - V^{\rm{kink}}_-| 
\lesssim  \lambda_- \delta \phi
\ll | V^{\rm kink}_+ -  V^{\rm kink}_-| \, .
\ee
Here, $\delta \phi$ is the scale on which the kink is resolved in the
smooth and regular potential, $\lambda_-$ denotes the slope in the
kink potential close to the true minimum and $V_{\mp}$ denote values
of the potential in the true (false) vacuum, respectively.

The first inequality reflects the fact that the bounce action varies
strongly if the field can accumulate a sizable kinetic energy in the
cap; the second inequality results from the fact that the bounce
action is very sensitive to the potential difference between the true
and the false vacuum in the thin-wall regime.

For example, this includes potentials that fluctuate strongly or that
do not have a finite second derivative in the true vacuum, as for
potentials that behave close to the true vacuum as $V \simeq (\phi -
\phi_-)^\alpha$ with $1<\alpha<2$. In particular, for the latter class
of potentials, the kink approximation yields accurate results even
though the bounce solutions cannot be obtained from the regular potential
using the conventional shooting algorithm.

Violations of the condition Eq.~\eqref{capcond} can appear for
instance within the context of the 4D effective field theory
approximation to a theory of quantum gravity such as string theory.
It is not clear how to self-consistently describe caps with curvature
larger than $M_p^2$ within this framework. In particular, in such a
high curvature cap, steep local maxima acting as large positive spikes
as discussed before cannot be excluded. In a situation where the cap
is confined to such a region of strong quantum gravity effects, we can
no longer guarantee that condition \eqref{capcond} is satisfied from a
calculation of the cap from effective field theory. Thus, a
description of the tunneling using the description of
\cite{Coleman:1977py} and \cite{Coleman:1980aw} entirely within the
realm of effective field theory is not possible for a high curvature
cap. In this case, full quantum gravity effects must be incorporated
to calculate the tunneling amplitude.

The kink approximation in terms of piecewise linear potentials~\cite{Duncan:1992ai} was used to estimate the tunneling rate in highly asymmetrical potentials far away from the thin-wall limit. Examples arise in theories of meta-stable dynamical supersymmetry breaking in ${\cal N}=1$ gauge theories~\cite{Intriligator:2006dd}, or their embedding into heterotic string theory~\cite{Serone:2007sv}. The kink approximation was used there without any prior justification of its validity, and might have led to qualitatively different life times, which motivated our work.

Beyond field theory, possible examples in string theory are e.g.~warped compactifications
with D3-branes. In such compactifications, the warp factor contributes
to the 4D effective scalar potential. To leading order, the warp
factor, and in turn its contribution to the 4D scalar potential, may
develop 'kinks' at the position of a D3-brane, similar to the case of
5D warped Randall-Sundrum compactifications~\cite{Randall:1999ee}. String theory effects
can smooth such 'kinks', but the curvature of the smoothing cap may
then be too large, as mentioned above, for a treatment within effective field theory alone.

\section*{Acknowledgments}
The authors wish to thank L.~Covi for stimulating discussions and
S.~Bobrovskyi for bringing the work \cite{Norton:2008} to our
attention. This work was supported by the Impuls und Vernetzungsfond
of the Helmholtz Association of German Research Centers under grant
HZ-NG-603, and German Science Foundation (DFG) within the
Collaborative Research Center 676 ``Particles, Strings and the Early
Universe''.

\bibliographystyle{kp}
\bibliography{linear_bounce}
\end{document}